\documentclass{PoS}

\title{Measurements of jet fragmentation and jet substructure with ALICE}

\ShortTitle{Jet fragmentation and jet substructure with ALICE}

\author{\speaker{Markus Fasel for the ALICE Collaboration}\\
        Oak Ridge National Laboratory, Oak Ridge, Tennessee, United States\\
        E-mail: \email{markus.fasel@cern.ch}}

\abstract{We discuss the latest results from jet fragmentation 
and jet substructure measurements performed with the ALICE experiment in
proton-proton and heavy-ion collisions in a wide range of jet transverse 
momentum. The jet production cross sections and cross section 
ratios for different jet resolution parameters will be shown in 
a wide range of $p_{\textrm{T}}$. Results will be compared to 
next-to-leading order pQCD calculations.}

\FullConference{7th Annual Conference on Large Hadron Collider Physics - LHCP2019\\
		20-25 May, 2019\\
		Puebla, Mexico}

\begin{document}

\section{Introduction}

Jet substructure measurements at low transverse momentum $p_{\textrm{T}}$ allow 
to constrain various perturbative and non-perturbative effects 
contributing to jet production, such as hadronization, 
perturbative radiation and the underlying event. For example the 
measurement of the jet cross section for various jet resolution 
parameters $R$ and the ratios of cross sections, in particular 
for $p_{\textrm{T}} < 100~\textrm{GeV}/c$ show different 
sensitivity to the various effects  \cite{Dasgupta:2016bnd}. In 
addition, the large datasets obtained with LHC Run2 allow for more 
differential studies. This is used in the measurement of the 
groomed momentum fraction $z_{\rm{g}}$ using the SoftDrop algorithm 
\cite{Larkoski:2014wba,Butterworth:2008iy}, defined as

\begin{equation}
    z_{\rm{g}} = \frac{min(p_{\rm{T},1}, p_{\rm{T}, 2})} {p_{\rm{T},1} + p_{\rm{T},2}} > z_{cut} \theta^{\beta}
\end{equation}

\noindent with $p_{\rm{T},1}$ and $p_{\rm{T},2}$ being the $p_{\rm{T}}$ 
of the hardest and second hardest subjet. The level of grooming 
is controlled by the parameters $z_{cut}$ and $\beta$. The observable 
is selected as it is closely related to the QCD splitting function 
at leading order \cite{Dasgupta:2016bnd}. No $p_{\textrm{T}}$-dependence 
is expected. The difference between the parton and the reconstructed 
jet momentum is expected to increase as log($R$) for perturbative 
radiation, to decrease as 1/$R$ due to hadronization effects and to 
increase as $\sim R^{2}$ due to the contribution from the underlying event 
\cite{Dasgupta:2016bnd}. In heavy-ion collisions the groomed momentum 
fraction is sensitive to effects from the hot and dense medium.

Jets are reconstructed using the anti-$k_{\textrm{T}}$ algorithm 
\cite{Cacciari:2008gp} from the FastJet package \cite{Cacciari:2011ma} 
employing the E scheme recombination. Clustering is done for track-based 
jets using tracks reconstructed in the ALICE tracking detectors, 
requiring a minimum constituent $p_{\textrm{T}}$ of 150 MeV/$c$. For fully 
reconstructed jets clusters in the electromagnetic calorimeter with 
energy larger than 300 MeV are included in the jet reconstruction.

The data used for the measurement in pp collisions at $\sqrt{s} 
= 13~\textrm{TeV}$ was collected in 2016 and 2017, consisting of 
a min. bias dataset with $L_{\textrm{int}} = 11.5~\textrm{nb}^{-1}$ 
and a jet-triggered dataset based on energy deposition in the 
electromagnetic calorimeter with $L_{\textrm{int}} = 4~\textrm{pb}^{-1}$. 
For the measurements in heavy-ion collisions the data was collected 
at $\sqrt{s_{\textrm{NN}}} = 2.76~\textrm{TeV}$ in 2011 and consists 
of 16M central collisions.

In heavy-ion collisions jet reconstruction is affected by the 
particles from the underlying event. Therefore observables with 
low sensitivity to the background, like jet-hadron correlations  
\cite{Adam:2015doa} are preferential. For substructure measurements 
the effect of the uncorrelated background can be handled in the 
detector response. For this PYTHIA events \cite{Sjostrand:2006za, 
Sjostrand:2014zea} are embedded into heavy-ion events, and the 
effect of the background is estimated and subtracted using area 
based \cite{Soyez:2012hv} or constituent based \cite{Berta:2014eza} 
subtraction methods.

\section{Measurement of the jet production in pp collisions at $\sqrt{s} = 13~\textrm{TeV}$}

\begin{figure}[ht]
    \centering
    \includegraphics[width=0.4\textwidth]{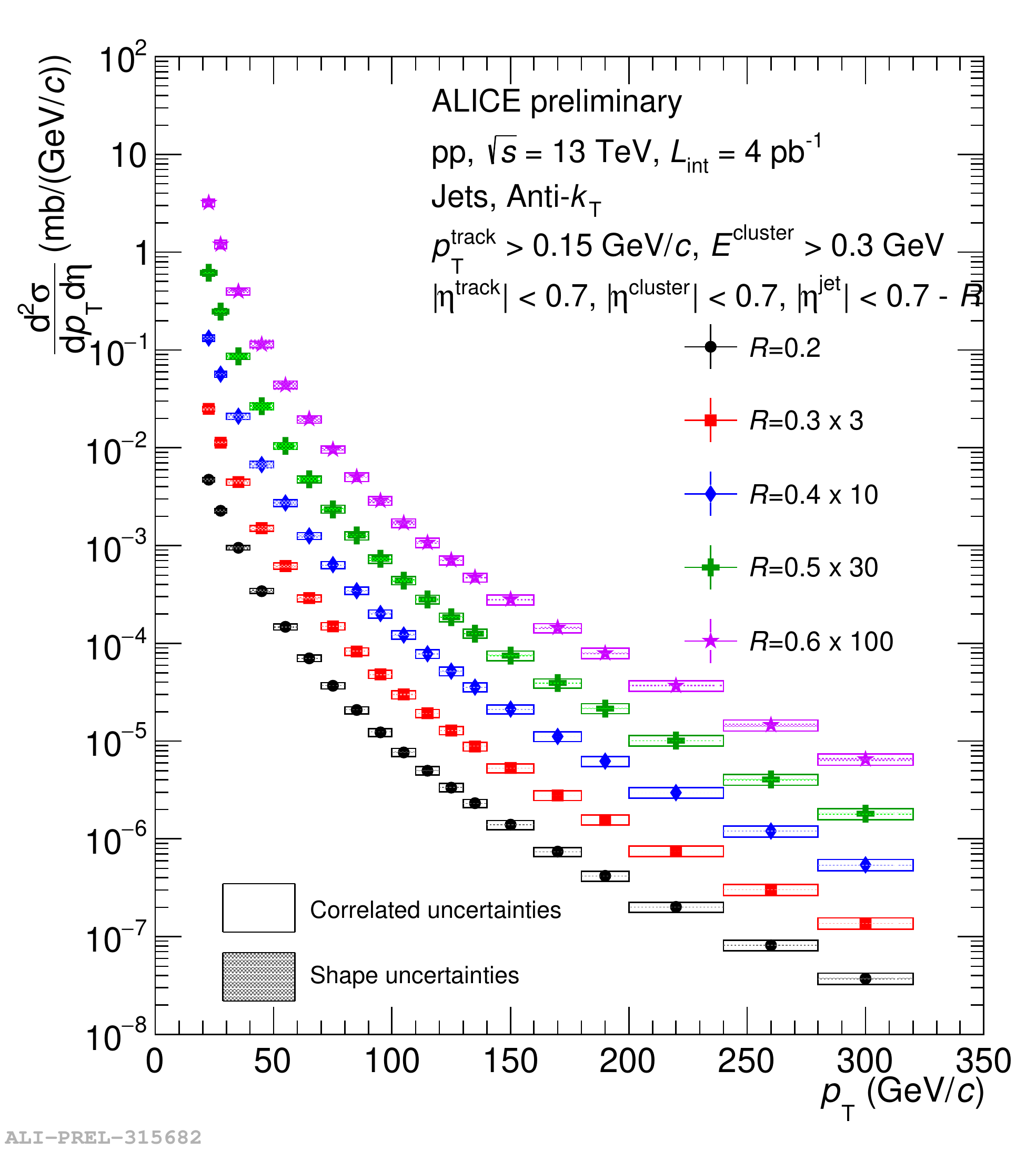}
    \includegraphics[width=0.4\textwidth]{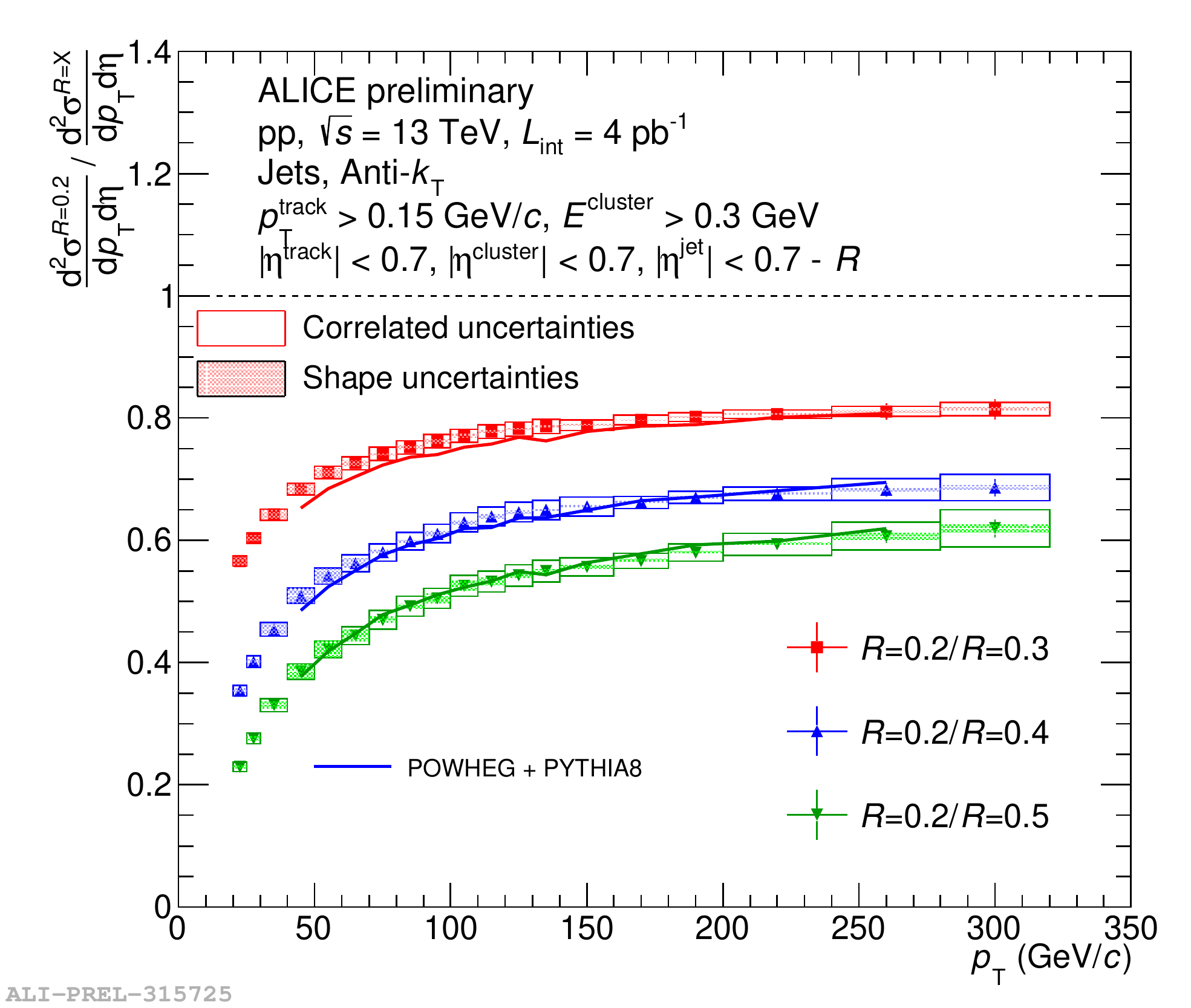}
    \caption{Left: $p_{\textrm{T}}$-differential jet production cross section in pp collisions
    at $\sqrt{\textrm{s}}$ = 13 TeV for different $R$. Right: Ratios 
    of jet production cross sections for jets with $R$ = 0.2 to the 
    cross section for various $R$. Lines indicate calculations 
    using POWHEG+PYTHIA6.}
    \label{fig:jetcrossection}
\end{figure}

\begin{figure}[ht]
    \centering
    \includegraphics[width=0.4\textwidth]{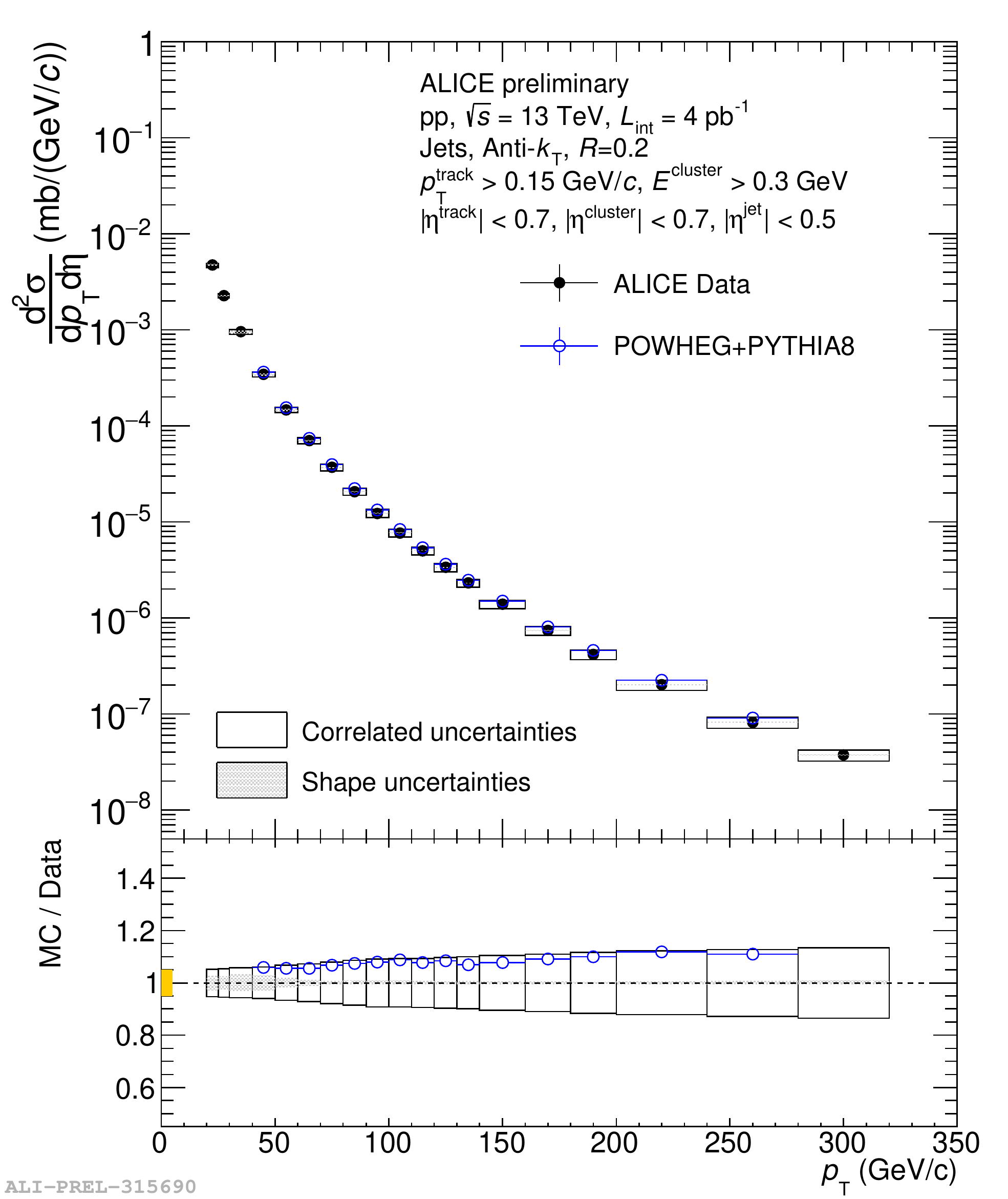}
    \includegraphics[width=0.4\textwidth]{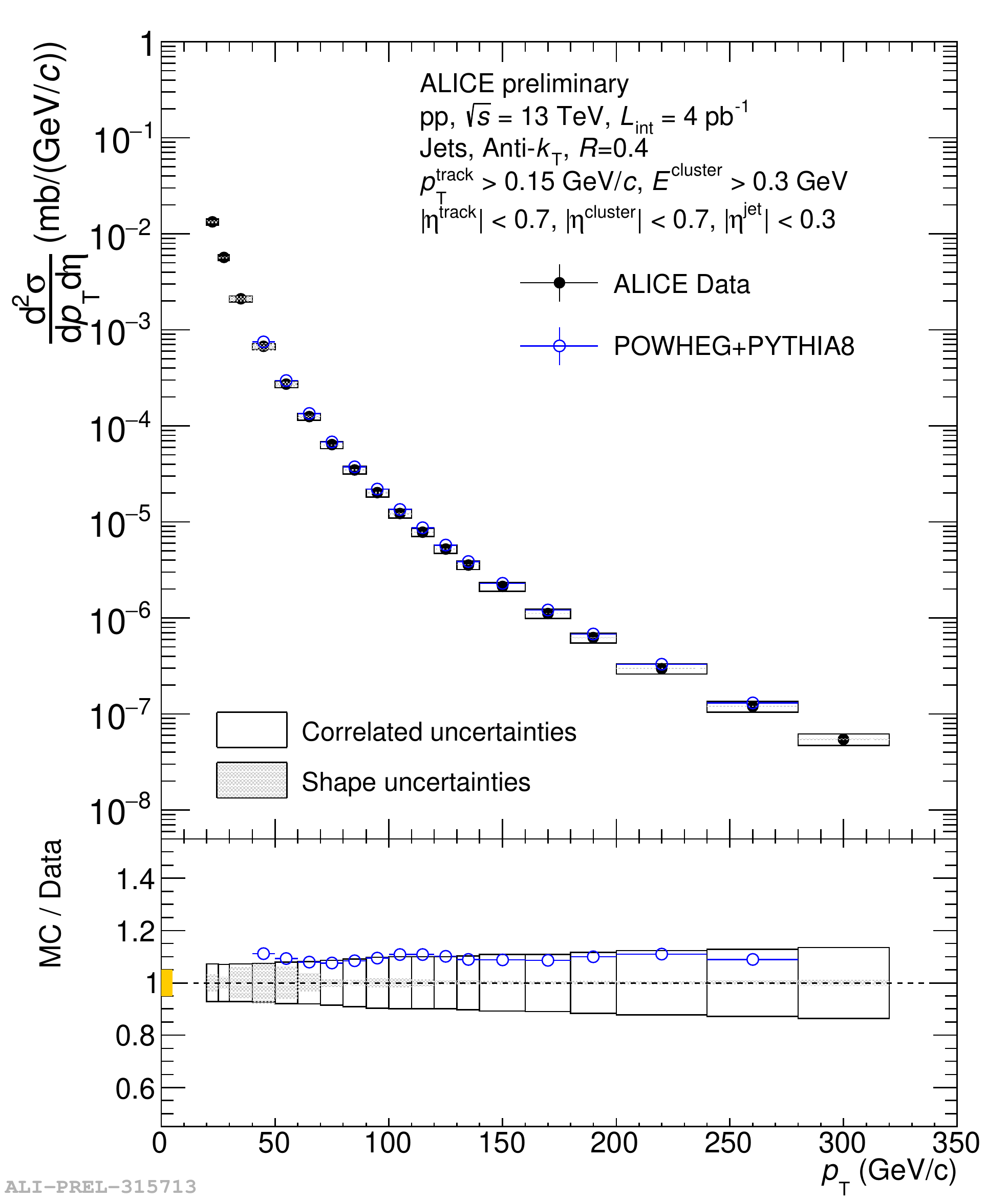}
    \caption{$p_{\textrm{T}}$-differential jet production cross section 
    for jets with $R$~=~0.2 (left) and $R$~=~0.4 (right) compared to 
    POWHEG+PYTHIA6 ($\mu_{R}$ = $\mu_{F}$ = 1).}
    \label{fig:crosssecPowheg}
\end{figure}

The $p_{\textrm{T}}$-differential cross section of jet production is 
measured for different jet radii between $R$ = 0.2 and $R$ = 0.6 for 
$20~\textrm{GeV}/c < p_{\textrm{T}} < 320~\textrm{GeV}/c$ 
(Fig.~\ref{fig:jetcrossection} (left)). The comparison to POWHEG+PYTHIA6 
\cite{Frixione:2007vw, Alioli:2010xd, Alioli:2010xa} is shown in 
Fig.~\ref{fig:crosssecPowheg} for $R$ = 0.2 (left) and $R$ = 0.4 (right). 
POWHEG calculations use $\mu_{R}$ = $\mu_{F}$ = 1. Calculations are in 
good agreement with the measured production cross sections. The ratios 
of the differential jet production cross section for jets  with $R$ = 
0.2 to the cross section for various $R$ is shown in 
Fig.~\ref{fig:jetcrossection} (right) for $R$ = 0.3 to $R$ = 0.6 in the same 
$p_{\textrm{T}}$ range. As many of the systematic uncertainties cancel 
in the ratios they can be measured with a higher precision. The 
cross section ratios are well-described by POWHEG+PYTHIA6. A good 
agreement with POWHEG+PYTHIA6 has also been found in cross section 
measurements of track-based jets and fully reconstructed jets  at 
lower centre-of-mass energies \cite{Mulligan:2018nnq, Acharya:2019tku,
Acharya:2018eat}.

\section{Measurement of the groomed momentum fraction $z_{g}$ in pp and heavy-ion collisions}

In pp collisions at $\sqrt{s}$ = 13 TeV the groomed momentum 
fraction $z_{\textrm{g}}$ has been measured for jets with 
$30~\textrm{GeV}/c < p_{\textrm{T}} < 200~\textrm{GeV}/c$ for various 
ranges in $p_{\textrm{T}}$. Jets were reclustered with the 
Cambridge/Aachen algorithm. The SoftDrop parameters 
$\beta$ = 0 and $z_{\textrm{g}}$ = 0.1 were used. No underlying event 
subtraction has been applied as the grooming already removes 
soft components from the underlying event. The distributions 
are unfolded back to particle level using a 2-dimensional 
unfolding based on the Bayesian method \cite{DAGOSTINI1995487}.
The response matrices have been obtained using a full detector
simulation based on GEANT3 \cite{Brun:1987ma} of pp events at the
same centre-of-mass energy generated with PYTHIA8 \cite{Sjostrand:2014zea}.

\begin{figure}
    \centering
    \includegraphics[width=0.3\textwidth]{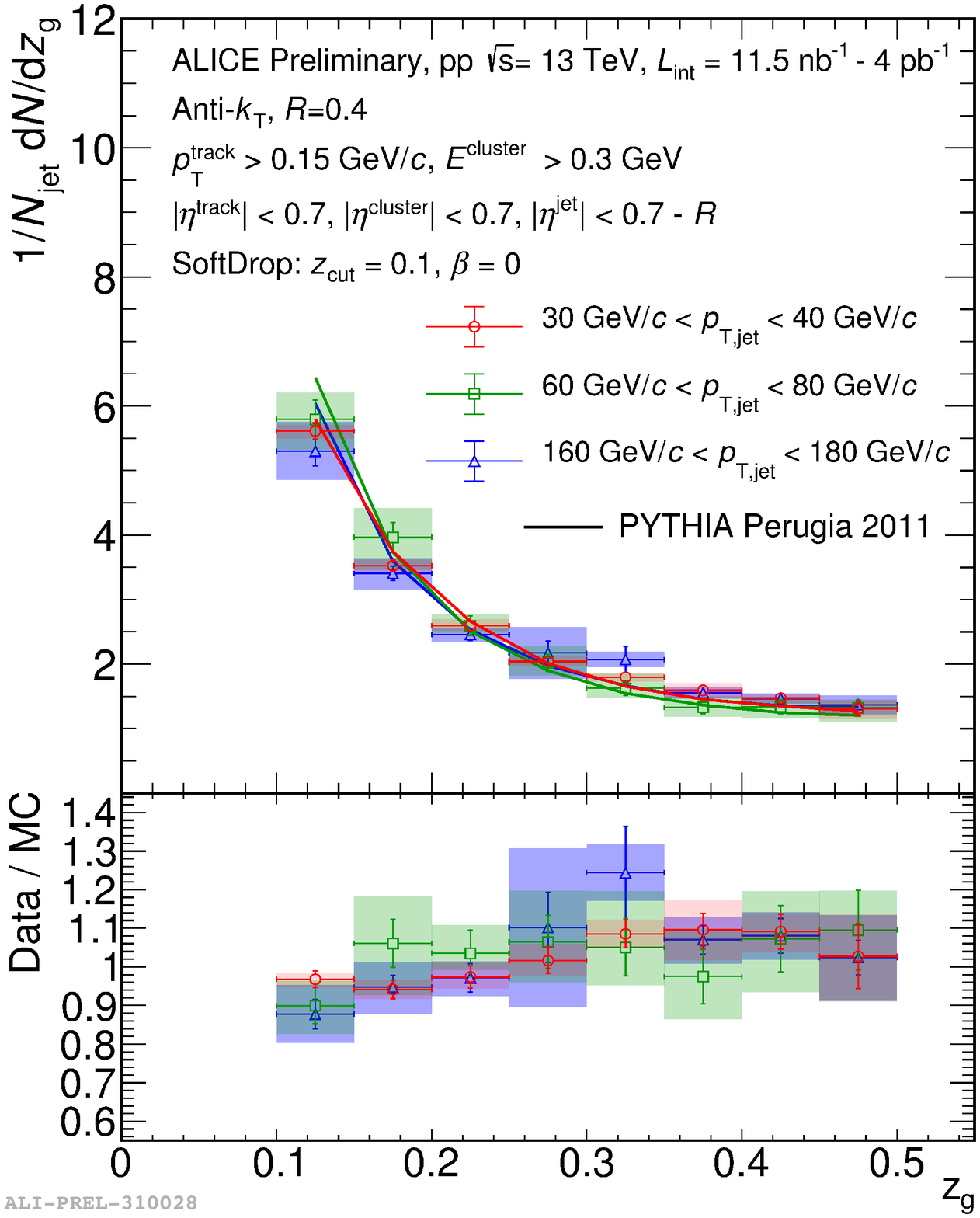}
    \includegraphics[width=0.3\textwidth]{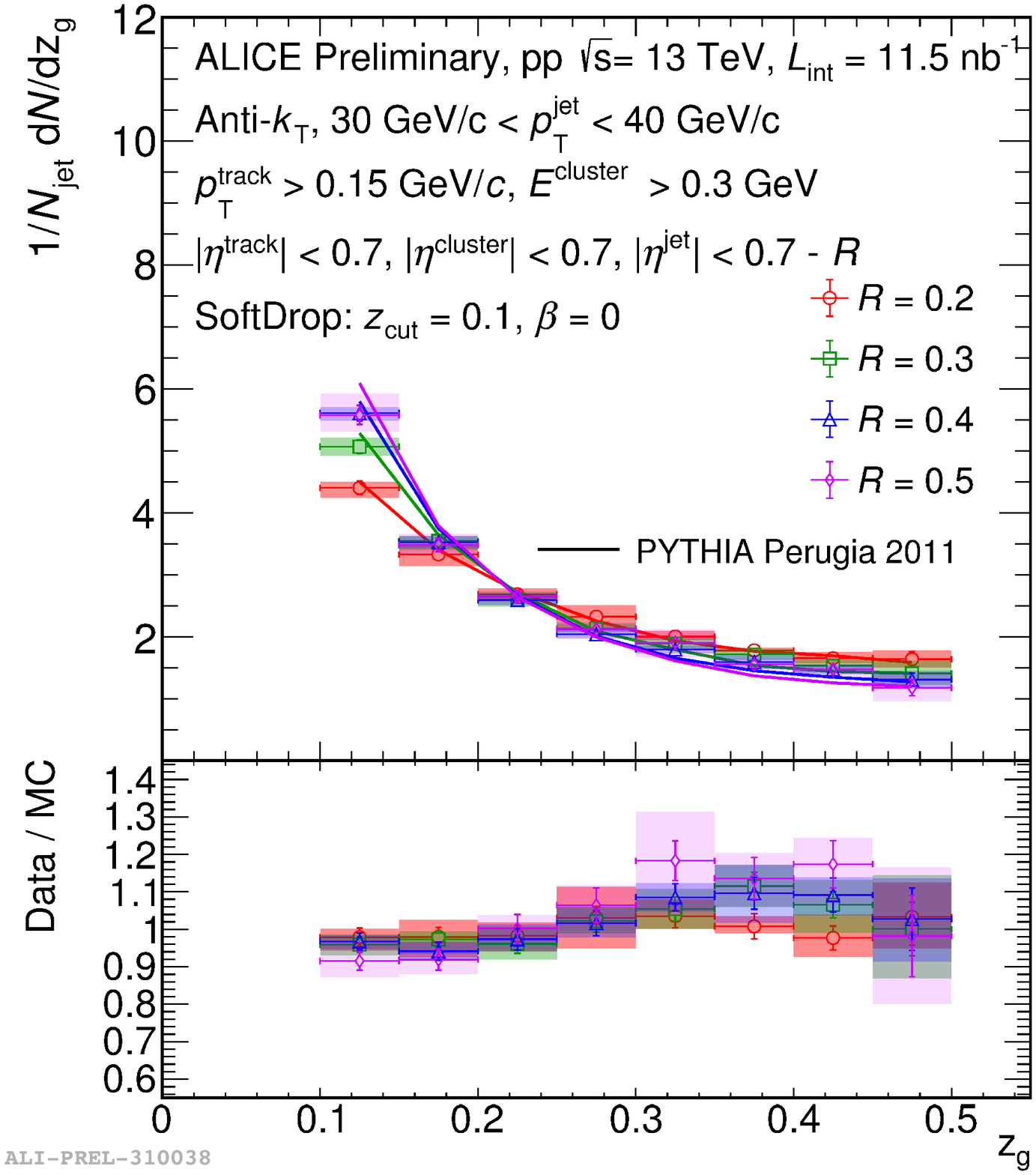}
    \includegraphics[width=0.3\textwidth]{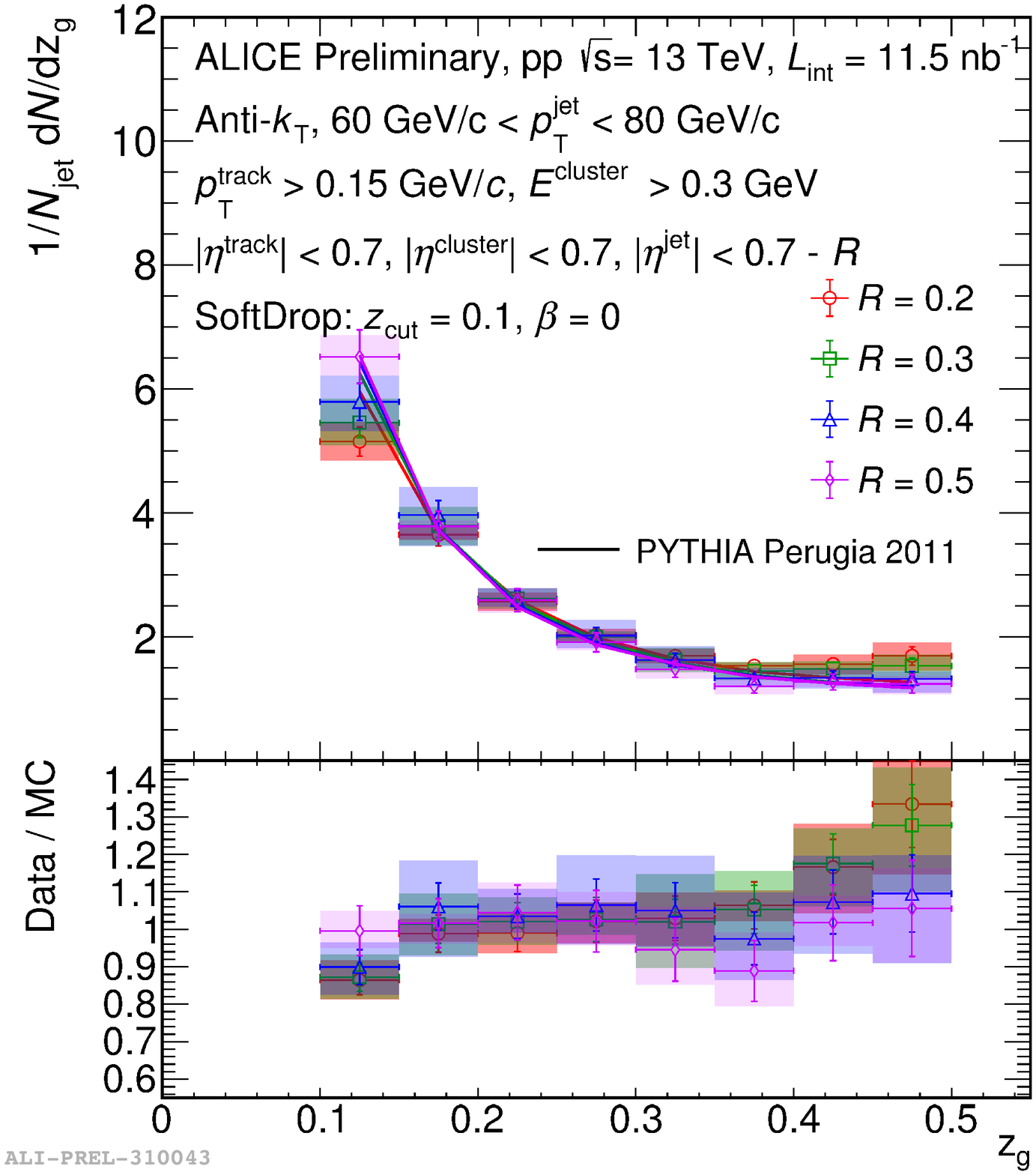}
    \caption{Left: $z_{\rm{g}}$-differential per-jet yield 
    for jets with $R$=0.4 for different bins in $p_{\textrm{T}}$ 
    (left), jets with different $R$ for $30~\textrm{GeV}/c < 
    p_{\textrm{T}} < 40~\textrm{GeV}/c$ (middle), and jets with 
    different $R$ for $60~\textrm{GeV}/c < p_{\textrm{T}} < 
    80~\textrm{GeV}/c$ (right). Lines indicate calculations 
    with the PYTHIA event generator using the Perugia 0 tune 
    \cite{Skands:2010ak}.}
    \label{fig:softdropfullpp}
\end{figure}

Fig.~\ref{fig:softdropfullpp} (left) shows the $z_{g}$-
differential jet yield for jets with R = 0.4 for 
$30~\textrm{GeV}/c < p_{\textrm{T}} < 40~\textrm{GeV}/c$,
$60~\textrm{GeV}/c < p_{\textrm{T}} < 80~\textrm{GeV}/c$,
and $160~\textrm{GeV}/c < p_{\textrm{T}} < 180~\textrm{GeV}/c$.
Results agree within the uncertainties, indicating no
$p_{\textrm{T}}$-dependence of the $z_{\textrm{g}}$
for jets with large resolution parameter. In the middle
and right panel the $z_{\textrm{g}}$-differential
yield of jets is plotted for different $R$ from $R~=~0.2$
to $R~=~0.4$ for jets with $ 30~\textrm{GeV}/c < 
p_{\textrm{T}} < 40~\textrm{GeV}/c$ (middle) and $60~
\textrm{GeV}/c < p_{\textrm{T}} < 80~\textrm{GeV}/c$ 
(right). For jets with lower $p_{\textrm{T}}$ a difference
in the $z_{\textrm{g}}$-differential yield for different
$R$ is observed, tending towards more symmetric splitting
for jets with smaller $R$. For jets with higher $p_{\textrm{T}}$
no dependence of the $z_{\textrm{g}}$ on $R$ is observed. The
trend is well reproduced by PYTHIA6 using the Perugia 0 tune.

\begin{figure}
    \centering
    \includegraphics[width=0.45\textwidth]{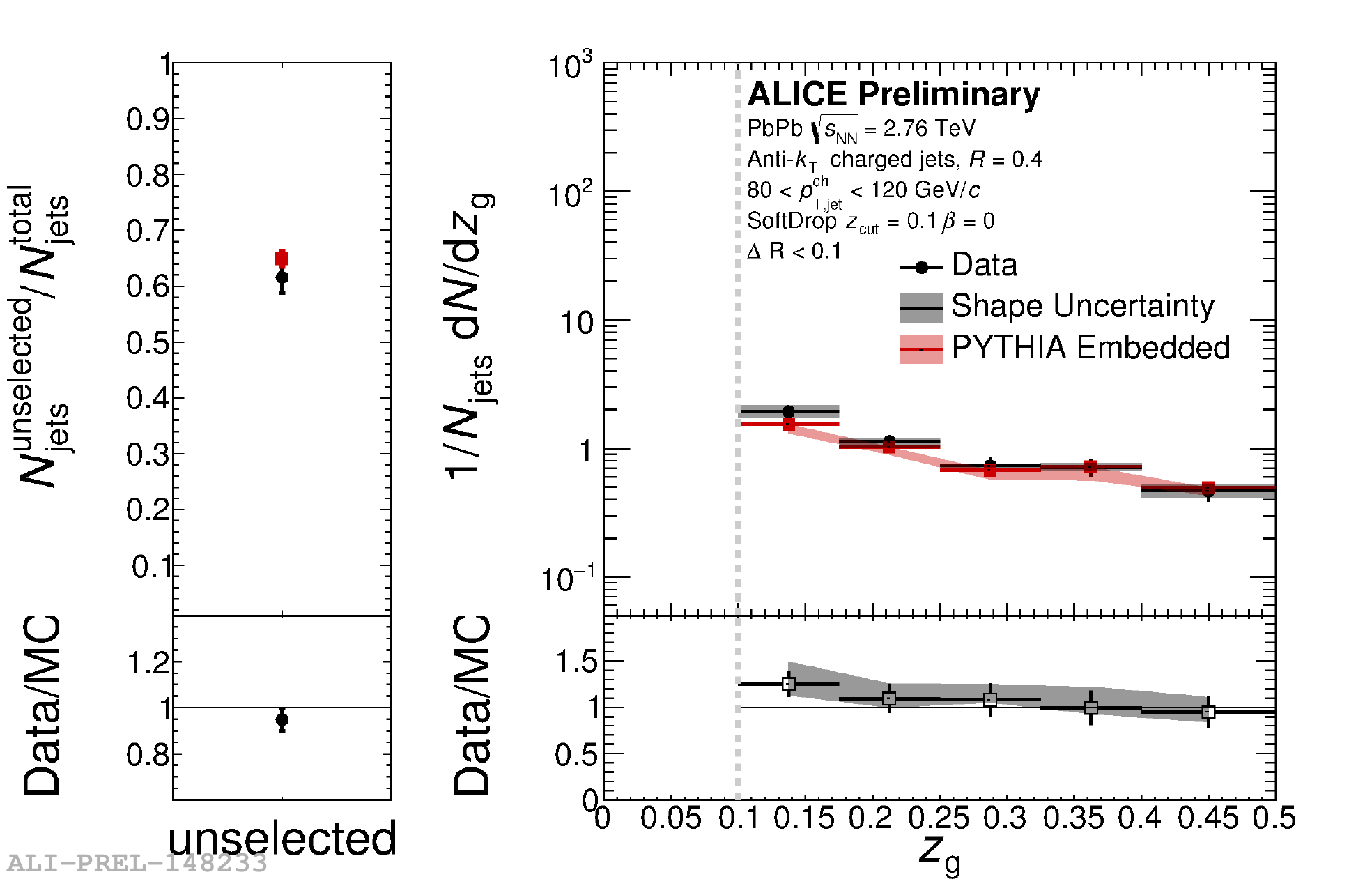}
    \includegraphics[width=0.45\textwidth]{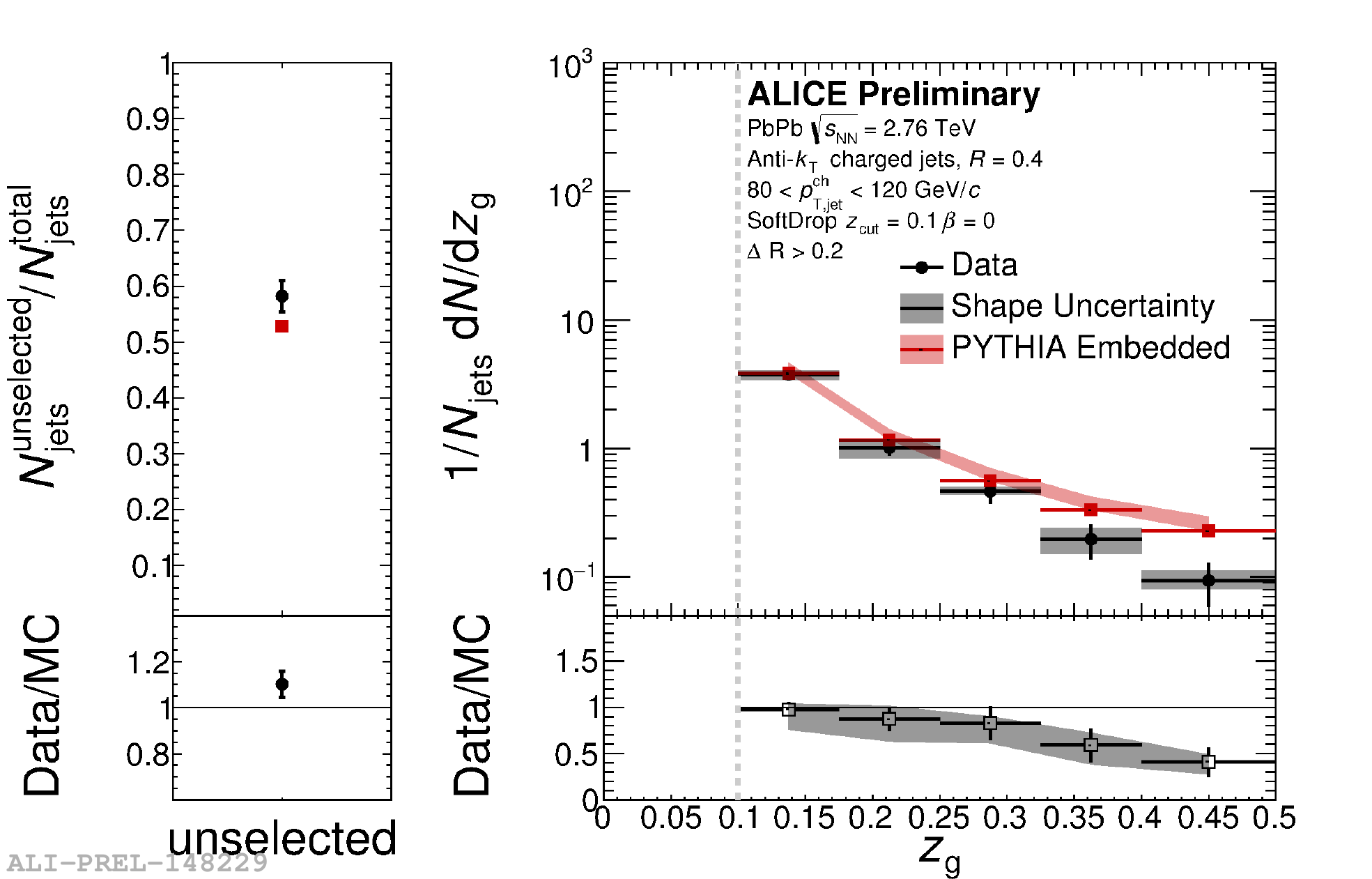}
    \caption{$z_{\textrm{g}}$-differential yield of
    track-based jets with $80~\textrm{GeV}/c < p_{\textrm{T}}
    < 120~\textrm{GeV}/c$ and $R~=~0.4$. \cite{Acharya:2019djg} 
    Left: jets with $\Delta R < 0.1$. Right: jets with 
    $\Delta\textrm{R} > 0.2$. Red bands indicating the distributions 
    from PYTHIA events embedded into heavy-ion collisions.}
    \label{fig:softdropchargedPbPb}
\end{figure}

In Pb-Pb collisions the $z_{\textrm{g}}$-differential yield
of track based jets was measured at $\sqrt{s_{\textrm{NN}}}$ 
= 2.76 TeV for jets with $R~=~0.4$ and $80~\textrm{GeV}/c < 
p_{\textrm{T}} < 120~\textrm{GeV}/c$ \cite{Acharya:2019djg}. 
Fig. \ref{fig:softdropchargedPbPb} shows the $z_{\textrm{g}}$-
differential yields for strongly collimated jets with 
$\Delta R < 0.1$, where $\Delta R$ is the 
distance between the subjets, and jets with $\Delta \textrm{R} > 0.2$.
For the comparison to the vacuum PYTHIA events were embedded
into PbPb events in order to account for effects from the 
underlying background. While no modification can be observed
for very collimated subjets, for jets with larger 
$\Delta R$ a suppression of jets with symmetric 
splittings with an enhancement of untagged jets can be observed.

\section{Outlook}

The study of jet substructure can be generalized by 
varying the grooming conditions or looking at all 
splittings and constructing the Lund plane 
\cite{Andersson:1988gp}. The Lund plane is obtained
directly from the QCD kernel. Recent studies 
\cite{Andrews:2018jcm} show that different regions
in the Lund plane are sensitive to different effects, among
them medium-induced radiation and coherent radiation. A 
precise measurement of the Lund plane in heavy-ion collisions
will allow to provide further constraints on the different
effects contributing to jet modification in the hot and dense
medium.

\bibliographystyle{JHEP}
\bibliography{biblio}

\end{document}